\documentclass[doublecol]{epl2}
\usepackage{graphicx}
\usepackage{amsmath}
\usepackage{amssymb}
\usepackage{color}
\usepackage{bm}

\newcommand{\beq}{\begin{equation}}
\newcommand{\eeq}{\end{equation}}
\newcommand{\beqa}{\begin{eqnarray}}
\newcommand{\eeqa}{\end{eqnarray}}

\title{Substrate rigidity deforms and polarizes active gels}

\author{S. Banerjee\inst{1} \and M.C. Marchetti\inst{1,2}}

\institute{
  \inst{1} Department of Physics, Syracuse University, Syracuse, New York 13244-1130, USA\\
  \inst{2} Syracuse Biomaterials Institute, Syracuse University, Syracuse, New York 13244-1130, USA
}

\abstract{We present a continuum model of the coupling between cells and substrate that accounts for some of the observed substrate-stiffness dependence of cell properties. The cell  is modeled as an elastic active gel, adapting  recently developed  continuum theories of active viscoelastic fluids. The coupling to the substrate enters as a boundary condition that relates the cell's deformation field to local stress gradients. In the presence of activity, the  coupling to the substrate yields spatially inhomogeneous contractile stresses and deformations in the cell and can enhance polarization, breaking the cell's front-rear symmetry. }

\begin{document}

\maketitle
\section{Introduction}
Many cell properties, including cell shape, migration and differentiation, are critically controlled by the strength and nature of the cell's adhesion to a solid substrate and by the substrate's mechanical properties~\cite{Discher2005}. For instance, it has been demonstrated that cell differentiation is optimized in a narrow range of matrix rigidity~\cite{Engler2004} and that the stiffness of
 the substrate can direct lineage specification of human mesenchymal stem cells~\cite{Engler2006}. In endothelial cells, adhesion to a substrate  plays a crucial role in guiding cell migration and controlling  a number of physiological
processes, including vascular development, wound healing, and tumor spreading~\cite{Reinhardt-King2008}.
Fibroblasts and endothelial cells seem to generate more traction force and develop a broader
and flatter morphology on stiff substrates than they do on soft but equally adhesive surfaces~\cite{Lo2000,Yeung2005}. They show an abrupt change in their spread area within a narrow
 range of substrate stiffnesses. This spreading also coincides with the appearance of stress fibers in the cytoskeleton, corresponding to the onset of a substantial amount of polarization within the cell~\cite{Yeung2005}.  Finally, such cells
preferentially move from a soft to a hard surface and migrate
faster on stiffer substrates~\cite{Guo2006}.
The mechanical interaction of cells with a surrounding matrix is to a great extent controlled by contractile forces  generated  by interactions between filamentary actin and myosin proteins in the cytoskeleton. Such forces are then transmitted by cells to  their surroundings through the action of focal adhesions that produce elastic stresses both in the cell and in the surrounding matrix. Cells in turn are capable of responding to the substrate stiffness by adjusting their own adhesion and elastic properties, with important implications for cell motility and shape~\cite{Discher2005,Barnhart2011}.

 In this letter we present a simple model of the coupling between cells and substrate that accounts for some of the observed substrate-stiffness dependence of cell properties. The cell itself is modeled as an elastic active gel, adapting  recently developed  continuum theories of active viscoelastic fluids~\cite{Kruse2004,Kruse2005,JKPJPhysRep2007}. In these models  the  transduction of chemical energy from ATP hydrolysis into mechanical work by myosin motor proteins pulling on actin filaments yields active contractile contributions to the local stresses.  The continuum theory of such  \emph{active liquids}  has led to several predictions, including the onset of
spontaneous deformation and flow in  active
films~\cite{Voituriez06,Giomi2008} and the retrograde flow of actin in the lamellipodium of crawling cells~\cite{JKPJPhysRep2007}. Active liquids
cannot, however, support elastic stresses at long times, as required
for the understanding of the crawling dynamics
of the lamellipodium and of active contractions in living cells.
Models of \emph{active elastic solids} on the other hand have been shown to account for the contractility and stiffening of in-vitro actomyosin networks
~\cite{Mizuno2007,MacKintoshLevine2008,LMJP-EPL2009} and the spontaneous oscillations of muscle sarcomeres ~\cite{Kruse2007,SBMCM2011}.  Very recently a continuum model of a one-dimensional polar, active elastic solid has also been used to describe the alternating polarity patterns observed in stress fibers~\cite{Marcq2010}.   In all these
cases the elastic nature of the network at low frequency is crucial
to provide the restoring forces needed to support  deformations and oscillatory
behavior.

\begin{figure}
\begin{center}
         \label{stress-and-u}
\includegraphics[width=0.3\textwidth]{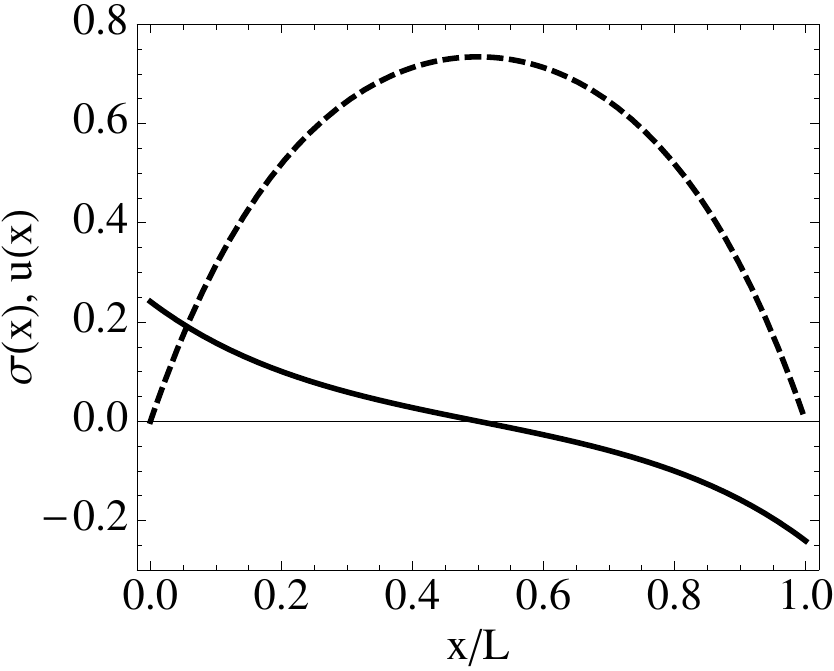}\\
          \label{dl}
\includegraphics[width=0.28\textwidth]{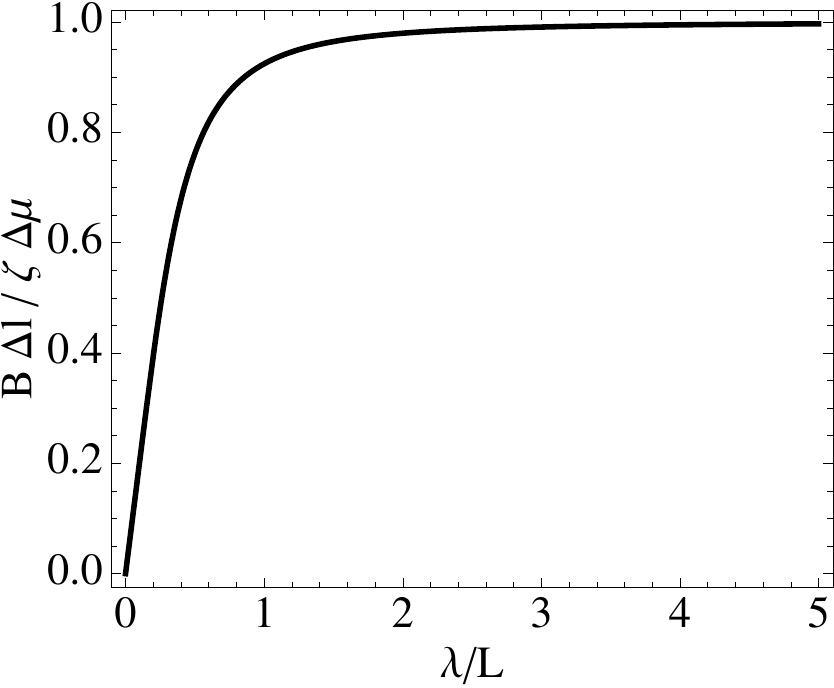}
\end{center}
\caption{Top: stress $\sigma(x)/\zeta\Delta\mu$ (dashed line) and deformation $u(x)B/\zeta\Delta\mu$ (solid line) profiles as functions of the position $x$ inside a cell of length $L$  for $\lambda/L=0.25$. Bottom: the cell's total deformation  $\Delta\ell=u(0)-u(L)$  as a function of $\lambda/L$. In the plot the deformation $\Delta\ell$ is normalized to its maximum value $\zeta\Delta\mu/B$.}
\label{isotropic}
\end{figure}

We model a cell as an elastic active film anchored to a solid substrate and study the static response of the film to variations in the strength of the anchoring. Although in the following we refer to our system as a cell, we stress that, on different length scales, the active elastic gel could also serve as a model for a confluent cell monolayer on a substrate. The coupling of the cell to the substrate enters via a boundary condition controlled by a ``stiffness" parameter  that depends on both the cell/substrate adhesion as well as the substrate rigidity. The description is macroscopic and applies on length scales large compared to the typical mesh size of the actin network in the cell lamellipodium (or large compared to the typical cell size in the case of a cell monolayer).  By solving the elasticity and force balance equations in a simple one-dimensional geometry we obtain several experimentally relevant results. First, in an isotropic active gel substrate anchoring yields  stresses and contractile deformations. The stress and deformation profiles for an isotropic active elastic gel are shown in the top frame of Fig.~\ref{isotropic}. The stress is largest at the center of the cell. Interestingly, a very similar  profile of \emph{tensile} stresses has been observed in confluent monolayers of migrating epithelial cells~\cite{Trepat2009}, where the stress increases as a function of the distance from the leading edge of the migrating layer and reaches its maximum at the center of the cell colony. Although our model considers stationary  active elastic layers (and the resulting stresses are contractile as opposed to tensile), in both cases these stresses originate from active processes in the cell, driven by ATP consumption. The deformation of the active layer is largest at the cell boundaries (see Fig.\ref{isotropic}, top frame), as seen in experiments imaging traction forces exerted by cells on substrates~\cite{Danuser2010} and its overall magnitude increases with cell activity. The density of the active gel layer is concentrated at the boundary, where the  local contractile deformations are largest. The net deformation of the cell over its  length is shown in the bottom frame of Fig.~\ref{isotropic} and it  increases monotonically with decreasing substrate stiffness,  in qualitative agreement with experiments on fibroblasts showing that these cells are more extended on stiff substrates~\cite{Yeung2005}. Finally,  if the cell is polarized on average, the coupling to the substrate  generates a spatially  inhomogeneous polarization profile inside the cell. The mean polarization is enhanced over its value in the absence of substrate anchoring and it is a non-monotonic function of substrate stiffness (see Fig.~\ref{polarization}). This result is in qualitative agreement with recent experiments that have demonstrated an intimate relation between the matrix rigidity and the alignment of cell fibers within the cell, suggesting that maximum alignment may be obtained for an optimal value of the substrate rigidity~\cite{Zemel2010}.

\section{The active gel model} The cell is modeled as an active gel described in terms of a  density, $\rho({\bf r},t)$, and a displacement field, ${\bf u}({\bf r},t)$, characterizing local deformations.   In addition, to account for the possibility of cell polarization as may be induced by directed myosin motion and/or filament treadmilling, we introduce a polar orientational order parameter field,  ${\bf P}({\bf r},t)$. Although we are describing a system out of equilibrium, it is convenient to formulate the model in terms of a local free energy density $f=f_{el}+f_{P}+f_{w}$,  with
 \begin{subequations}
 \begin{gather}
 \label{f-el}
 f_{el}= \frac{B}{2}u_{kk}^2+G\tilde{u}_{ij}^2\;,\\
 \label{f-P}f_P=\frac{a}{2}|{\bf P}|^2+\frac{b}{4}|{\bf P}|^4+\frac{K}{2}(\partial_iP_j)(\partial_jP_i)\;,\\
 \label{f-w} f_w=\frac{w}{2}(\partial_iP_j+\partial_jP_i)u_{ij}+w'(\bm\nabla\cdot{\bf P})u_{kk}\;,
 \end{gather}
 \end{subequations}
Here $f_{el}$ is the energy of elastic deformations, with $B$ and $G$ the compressional and shear elastic moduli of the gel, respectively,  $u_{ij}=\frac12(\partial_iu_j+\partial_ju_i)$  the  symmetrized strain tensor, with $\tilde{u}_{ij}=u_{ij}-\frac{1}{d}\delta_{ij}u_{kk}$ and $d$ the dimensionality. The first two terms in Eq.~\eqref{f-P}, with $b>0$,  allow the onset of a homogeneous polarized state when $a<0$; the last term is the energy cost for spatially inhomogeneous deformations of the polarization. We have used an isotropic elastic constant approximation, with $K$ a stiffness parameter characterizing the cost of both splay and bend deformations. Finally, the contribution $f_w$ couples strain and polarization and is unique to polar systems~\cite{Giomi2008,Marcq2010}. It describes the fact that in the active polar system considered here, like in liquid crystal elastomers, a local strain is always associated with a local gradient in polarization.  Such gradients will align or oppose each other depending on the sign of the phenomenological parameters $w$ and $w'$, which are controlled by microscopic physics. A positive sign indicates that an increase of density is  accompanied by positive splay (or enhanced polarization in one dimension). In active actomyosin systems filament polarity can be induced by both myosin motion and by treadmilling. If the polarization is defined as positive when pointing towards the plus (barbed) end of the filament, i.e., the direction towards which myosin proteins walk, the forces transmitted by myosin procession will yield filament motion in the direction of negative polarization, corresponding to $w<0$~\cite{TBLMCM2003}. In contrast, treadmilling, where polarization occurs at the barbed end, corresponds to $w>0$.  Density variations $\delta\rho=\rho-\rho_0$ from the equilibrium value, $\rho_0$, are slaved to the local strain according to $\delta\rho/\rho_0=-\bm\nabla\cdot {\bf u}$. The stress tensor is written as the sum of reversible and active contributions as $\sigma_{ij}=\sigma_{ij}^{r}+\sigma_{ij}^a$, where $\sigma_{ij}^r=\frac{\partial f}{\partial u_{ij}}$. The two contributions are given by
\begin{subequations}
\begin{gather}
\label{sigma-r}
 \sigma_{ij}^{r}
 =\delta_{ij}Bu_{kk}+2G\tilde{u}_{ij}+\frac{w}{2}\left(\partial_iP_j+\partial_jP_i\right)+w'\bm\nabla\cdot{\bf P}\delta_{ij}\;,\\
\label{sigma-a} \sigma_{ij}^a=\zeta(\rho)\Delta\mu\delta_{ij}+\zeta_\alpha\Delta\mu~ P_i P_j \;.
 \end{gather}
 \end{subequations}
Active stresses arise because the gel is driven out of equilibrium by continuous input of energy from the hydrolysis of ATP, characterized by the chemical potential difference $\Delta\mu$ between ATP and its products. For simplicity, we assume here $\Delta\mu$ to be constant, although situations where inhomogeneities in $\Delta\mu$ may arise, for instance, from inhomogeneous myosin distribution within the actin lamellipodium are also of interest.  The experimentally observed contractile effect of myosin corresponds to positive values of the coefficients $\zeta$ and $\zeta_\alpha$, that characterize the isotropic and anisotropic stress per unit $\Delta\mu$, respectively, due to the action of active myosin crosslinkers~\cite{Simha2002,Hatwalne2004,Kruse2005}. In polar gels there are also active stresses proportional to $\Delta\mu (\partial_i P_j + \partial_j P_i)$~\cite{Ahmadi2006,Giomi2008}. We neglect these terms here as terms of similar structure already arise from the coupling terms in $f_w$. By letting $\rho=\rho_0-\rho_0\bm\nabla\cdot{\bf u}$, we can write $\zeta(\rho)\Delta\mu\simeq \zeta_0\Delta\mu-\zeta_1\Delta\mu u_{kk}$. The second term describes active renormalization of the compressional modulus $B$ of the gel  and can yield a contractile instability~\cite{Kruse2007,SBMCM2011}. These effects have been described elsewhere~\cite{SBMCM2011} and will not be discussed here, where we will assume we are in a regime where the gel is elastically   stable. Finally, we note that the parameters $a$, $w$ and $w'$ may also in general depend on $\Delta\mu$ as cell polarity is induced by ATP-driven processes. For simplicity we keep these parameters fixed below.

Force balance requires
\beq
\label{force-balance}
\partial_j\sigma_{ij}=0\;.
\eeq
The coupling to the substrate (assumed for simplicity isotropic) is introduced as a boundary condition~\cite{KruseJoannyJulicherProst2006}  by requiring $\left[\sigma_{ij}\hat{n}_j\right]_{{\bf r}_s}=Eu_i({\bf r}={\bf r}_s)$, where $\hat{n}$ is a unit normal to the substrate and both sides of the equation are evaluated at points ${\bf r}_s$ on the substrate. Although in the following we will often refer to the parameter $E$ as the substrate stiffness, it is important to keep in mind that $E$  is controlled not only by the substrate rigidity, but also by the properties of the cell/substrate adhesions~\cite{Murray1984}.  Anisotropic substrates are not considered here, but can be described by a generalized boundary condition where $E$ is  a tensor quantity and will be discussed in a later publication. Finally, variations in the  polarization are described by the equation
 \beq
 \label{P}
 \partial_t{\bf P}+\beta\left({\bf P}
\cdot\bm\nabla\right) {\bf P}=\Gamma {\bf h}\;,
 \eeq
 with $\beta$ an advective coupling arising from ATP driven processes, such as treadmilling~\cite{Ahmadi2006,Giomi2008}, $\Gamma$ an inverse friction, and ${\bf h}=-\frac{\delta f}{\delta {\bf P}}$ the molecular field,
 given by
 \beq
 \label{h}
 h_i=-\left(a+b|{\bf P}|^2\right) P_i+K\nabla^2 P_i+w\partial_ju_{ij}+w'\partial_iu_{kk}\;.
 \eeq
Here $\beta$ is an active  velocity and is controlled by the activity $\Delta\mu$. In the following we write $\beta/(L\Gamma)=\zeta_\beta\Delta\mu$, with $L$ the typical size of the active gel.

 \section{Isotropic cell} We begin by considering the case of an isotropic  cell  and neglect the coupling to polarization. For simplicity, we consider a quasi-one-dimensional model where the cell is a thin sheet of active gel of thickness $z$ extending from $x=0$ to $x=L$, with $L>>h$. The substrate is flat and located at $z=0$. Although this is of course a gross simplification, we will see below that it captures the substrate-induced stresses and deformations and their dependence on substrate stiffness.  More realistic planar or thin film geometries  will be discussed in a future publication. Force balance yields $\partial_x\sigma_{xx}+\partial_z\sigma_{xz}=0$. Integrating over the thickness of the film, using $\sigma_{xz}(x,z=h)=0$ and  $\sigma_{xx}(x,z=0)=Eu_x(x,0)$, and letting
 $\sigma=\frac{1}{h}\int_0^h dz\sigma_{xx}(x,z)$, we obtain $\partial_x\sigma=E u_x(x,0)/h$.   In the limit $h<<L$, we  neglect all $z$ dependence  and assume that the only component of the displacement field is $u_x(x,0)\equiv u(x)$. Combining then the expression for the mean stress, $\sigma=B\partial_xu+\zeta_0\Delta\mu$  with the boundary condition we obtain
 \begin{equation}
 \label{sigm-eq-iso}
\sigma=\lambda^2\frac{d^2\sigma}{dx^2} + \zeta_0\Delta\mu\;,
\end{equation}
where $\lambda=\sqrt{B h/E}$ is a length scale controlled by the interplay of cell and substrate stiffness. The solution of this equation with boundary conditions $\sigma(x=0)=\sigma(x=L)=0$ is
\beq
\sigma(x)=\zeta\Delta\mu\left(1-\frac{\cosh{[(L-2x)/2\lambda]}}{\cosh{(L/2\lambda)}}\right)\;.
\eeq
The deformation field is then given by
\beq
u(x)=\frac{\zeta\Delta\mu \lambda}{B}\frac{\sinh{[(L-2x)/2\lambda]}}{\cosh{(L/2\lambda)}}\;.
\eeq
A finite activity $\Delta\mu$ generates stresses and deformations in the cell, as shown in the top frame of Fig.~\ref{isotropic}. In an isotropic gel, both the stress and the displacement profiles are symmetric about the cell's mid point and the cell is uniformly contracted. The deformation is localized near the cell's boundaries. The  length scale $\lambda$ determined by the ratio of cell to substrate stiffness controls the penetration of the deformation to the interior of the cell. If $\lambda\sim L$, corresponding to a substrate rigidity $E_L\sim B h/L^2$, the active stresses and deformation extend over the entire cell. For a cell layer of length $10\ \mu m$, thickness $1\ \mu m$ and elastic modulus $B\sim 100\ kPa$,  the substrate rigidity parameter $E_L$ can be estimated to be $\sim1\ kPa/\mu m$. The total deformation $\Delta\ell=u(0)-u(L)$ grows with activity and is shown in Fig.~\ref{isotropic} (bottom frame) as a function of $\lambda/L\sim1/\sqrt{E}$. The contraction decreases with increasing substrate stiffness  and saturates to a finite value for soft substrates.

\begin{figure}
\begin{center}
          \label{stress-asym}
\includegraphics[width=0.31\textwidth]{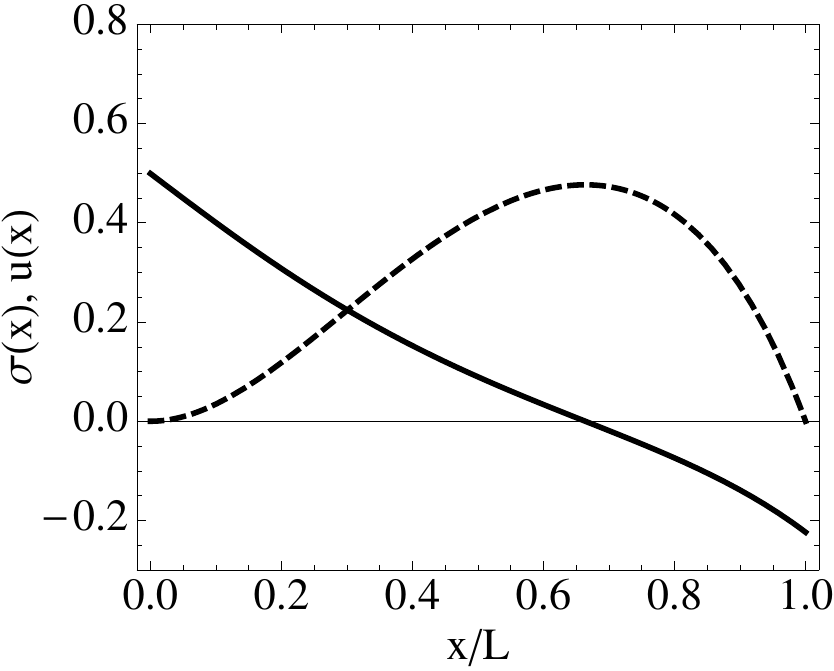}
\end{center}
\caption{ The stress $\sigma(x)/\zeta\Delta\mu$ (dashed line) and displacement $u(x) B/\zeta\Delta\mu$ (solid line) profiles of a cell on a substrate a constant stiffness gradient, described by  $E(x)=E_0x/L$ are shown as functions of the position $x$ inside the cell for   $\lambda/L=0.25$. The profiles are asymmetric and the stress is localized near $x=L$ where the stiffness is largest. }
\label{gradient}
\end{figure}

It is also interesting to consider a substrate of varying stiffness, as such substrates can be realized in experiments. We consider a constant stiffness gradient, corresponding to $E(x)=E_0x/L$. In this case Eq.~\eqref{sigm-eq-iso} becomes
\begin{equation}
\sigma=\frac{\lambda^2L}{x}\left( \frac{d^2\sigma}{dx^2} -   \frac{1}{x}\frac{d\sigma}{dx}  \right)+ \zeta_0\Delta\mu
\end{equation}
A closed solution can be obtained in terms of hypergeometric functions. The corresponding stress and displacement profiles are now asymmetric and are shown in   Fig.~\ref{gradient}. The stress is largest in the region of stiffest substrate, with a correspondingly smaller cell deformation. In other words, the largest cell deformation is obtained in the boundary region where the substrate is softest. In real cells the region where the substrate is softer and the resulting stresses in the cell are smaller may correspond to region of reduced focal adhesions.  Hence the gradient stiffness may yield a gradient in the strength of cell-substrate adhesion, providing a possible driving force for durotaxis, the tendency of cells to move from softer to stiffer regions~\cite{Lo2000,Wong2003,referee}.

 \section{Polarized cell}
We now consider the case of a polarized cell,  described by the full free energy $f$. The cell is modeled again as a thin film of length $L$ in the quasi-$1d$ geometry described earlier.  We are interested in steady state configurations. In the chosen geometry these are given by the solutions of the equations
\begin{subequations}
\begin{gather}
\label{boundary}\frac{d\sigma}{dx}=\frac{E}{h}u\\
\label{sigma-p}\sigma=B\frac{du}{dx} + \zeta_0\Delta\mu + \zeta_\alpha\Delta\mu\ p^2 + 2w\frac{dp}{dx} \\
\label{px}\zeta_\beta\Delta\mu L p\frac{dp}{dx}=K \frac{d^2p}{dx^2} +2w\frac{d^2u}{dx^2} -\left(a + bp^2\right)p
\end{gather}
\end{subequations}
where ${\bf P}=p(x){\bf \hat{x}}$ and  we have let $w'=w$ and $\beta/(L\Gamma)=\zeta_\beta\Delta\mu$. In the following we scale lengths with the cell's length $L$ and stresses with the cell's compressional modulus $B$. By combining Eqs.~\eqref{boundary}-\eqref{px}, we can eliminate $u$ and rewrite them as coupled equations for $\tilde\sigma=\sigma/B$ and $p$ as
\begin{subequations}
\begin{gather}
\label{sigma-p2}\tilde\sigma=\frac{\lambda^2}{L^2}\tilde\sigma'' + \nu_0 + \nu_\alpha p^2 + \tilde{w}p' \\
\label{px2}\left(\nu_{\beta}+2\nu_\alpha \tilde{w}\right) p p'=\tilde{K} p'' +\tilde{w} \tilde{\sigma}' -\left(\tilde{a} + \tilde{b}p^2\right)p
\end{gather}
\end{subequations}
where the prime denotes a derivative with respect to $x/L$, $\nu_{0,\alpha,\beta}=\zeta_{0,\alpha,\beta}\Delta\mu/B$,  $\tilde{w}=2w/BL$, $\tilde{a}=a/B$, $\tilde{b}=b/B$, and $\tilde{K}=K/(BL^2)-\tilde{w}$.
Thermodynamic stability requires $\tilde{K}>0$. As discussed in Ref.~\cite{Marcq2010} there could be possible active contributions to the coupling $w$, which at high activity leads to an alternating polarity pattern in the gel. Here we restrict ourselves to $\tilde{K}>0$.
\begin{figure}
\begin{center}
\includegraphics[width=0.35\textwidth]{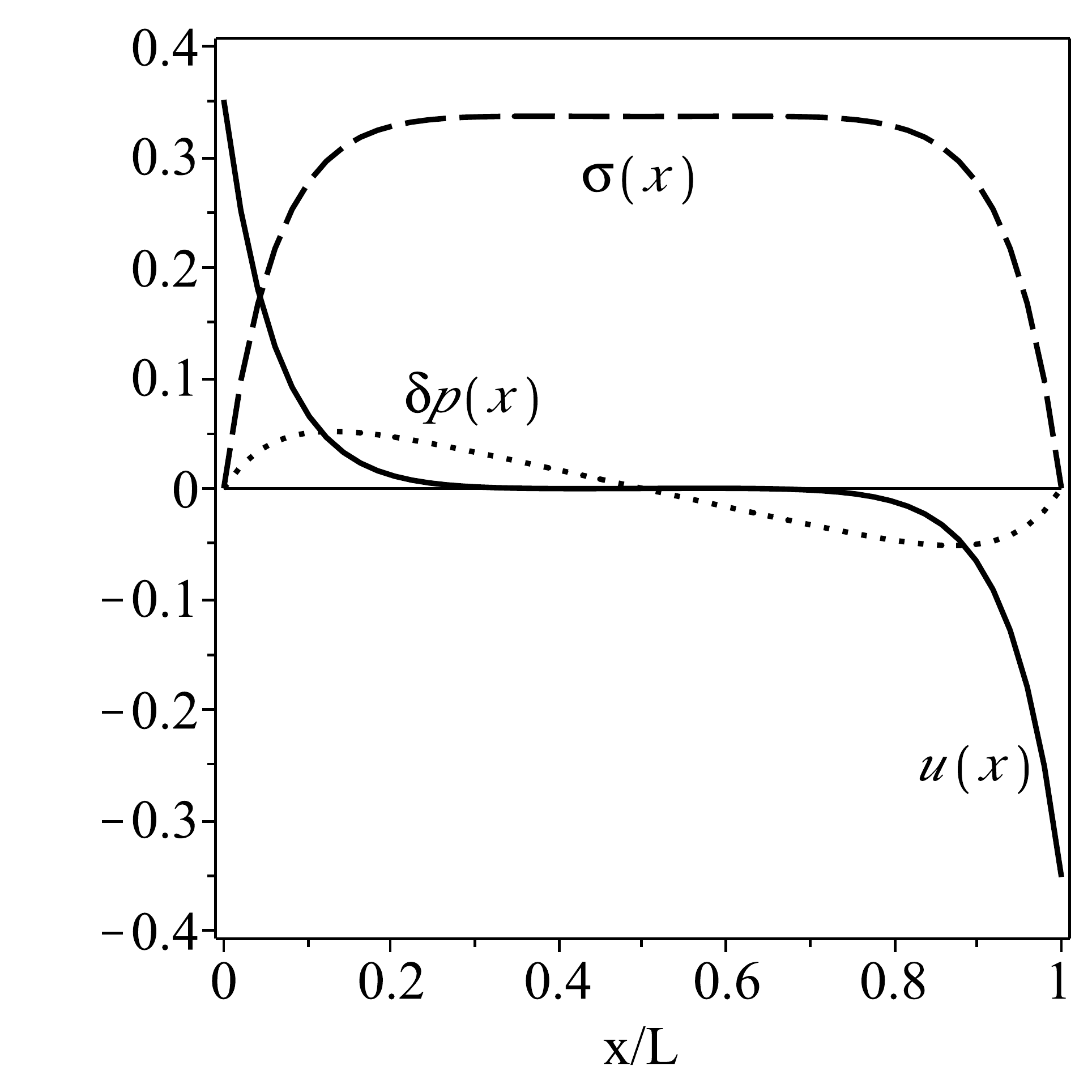}\\
\includegraphics[width=0.35\textwidth]{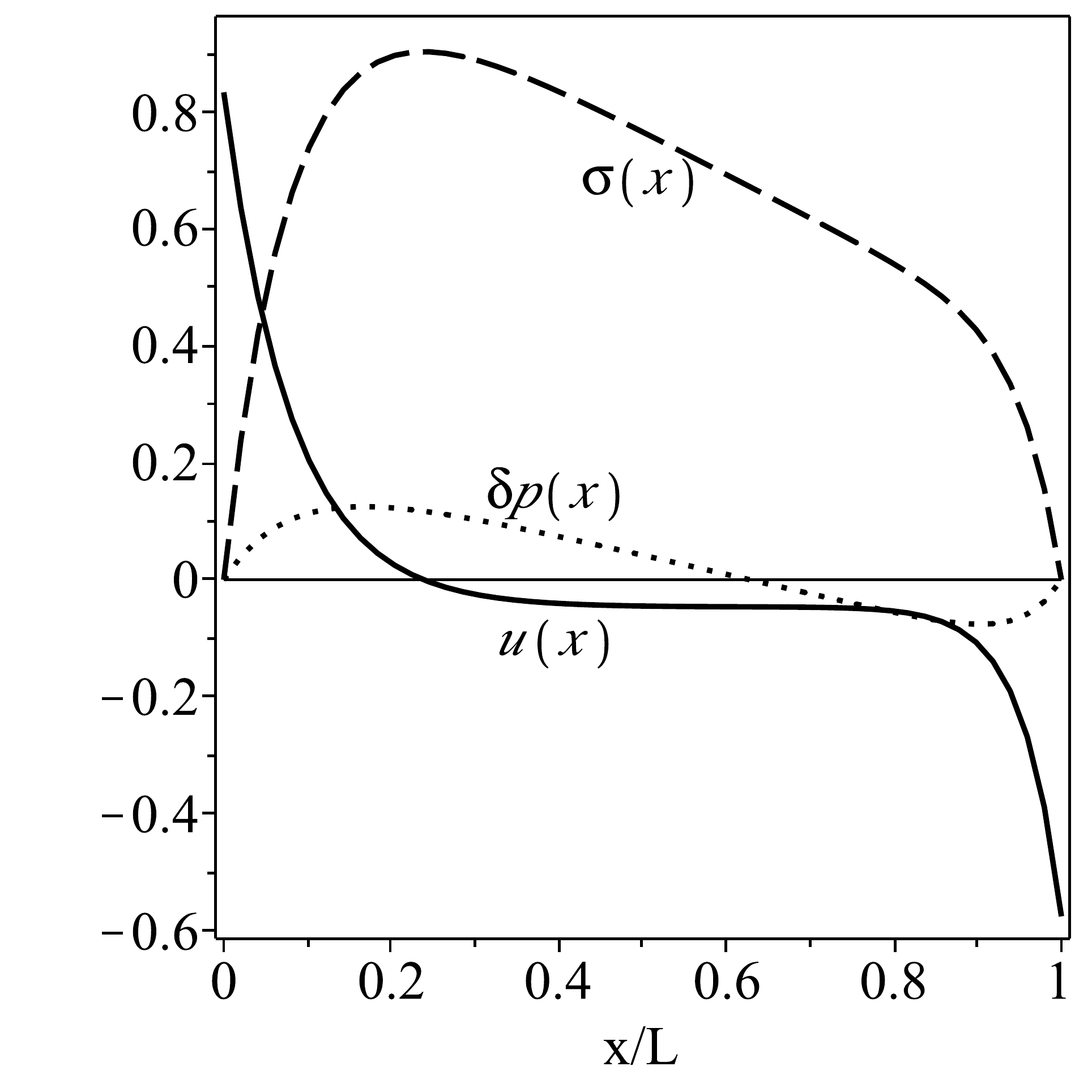}
\end{center}
\caption{Stress $\sigma(x)/B$ (dashed line), deformation field  $u(x)/L$ (solid line), and polarization   $\delta p(x)=p(x)-p_0$ (dotted line) profiles obtained by numerical solution of Eqs. \eqref{sigma-p2} and \eqref{px2} for two sets of boundary conditions on the polarization: $p(0)=p(L)=0$ (top frame) and $p(0)=p(L)=1$ (bottom frame).  Both plots are for $\lambda/L=0.25$, $\tilde{w}=4$,  $\nu_0=\nu_\alpha=\nu_\beta=1$, $\tilde{a}=\tilde{b}=1$, $\tilde{K}=1$. }
\label{profiles_p0}
\end{figure}

In the absence of activity ($\Delta\mu=0$) Eqs.~\eqref{sigma-p2} and \eqref{px2}  have two homogeneous solutions that satisfy the boundary condition $\sigma(0)=\sigma(L)=0$, corresponding to an isotropic state for $a>0$, with $p(x)=u(x)=0$ and to a polarized state for $a<0$, with $p(x)=p_0=\sqrt{-a/b}$ and $u(x)=0$. In both cases $\sigma(x)=0$.

For finite activity ($\Delta\mu \neq 0$), we find two qualitatively different solutions, depending on the boundary conditions used for the polarization. When Eqs.~ \eqref{sigma-p2} and \eqref{px2} are solved with  boundary condition $p(0)=p(L)=0$, consistent with an isotropic state in the limit $\Delta\mu=0$, the stress is an even function of $x$, as shown in the top frame of Fig.~\ref{profiles_p0}. It exhibits a maximum at $x=L/2$ and is symmetric about the mid point of the cell. Both the displacement and the polarization vanish at $x=L/2$ and are odd functions of $x$ about this point. For $a<0$ we solve the nonlinear equations with boundary condition $p(0)=p(L)=\sqrt{-a/b}$, consistent with a polarized state in the limit $\Delta\mu=0$. In this case the stress, deformation and polarization profiles are all asymmetric, as shown in the bottom frame of Fig.~\ref{profiles_p0}. The sign of the anisotropy is controlled by the sign of the polar coupling $w$. The figure displays the case  $w>0$, corresponding to filament convection towards the direction of positive polarization.

\begin{figure}
\begin{center}

\includegraphics[width=0.4\textwidth]{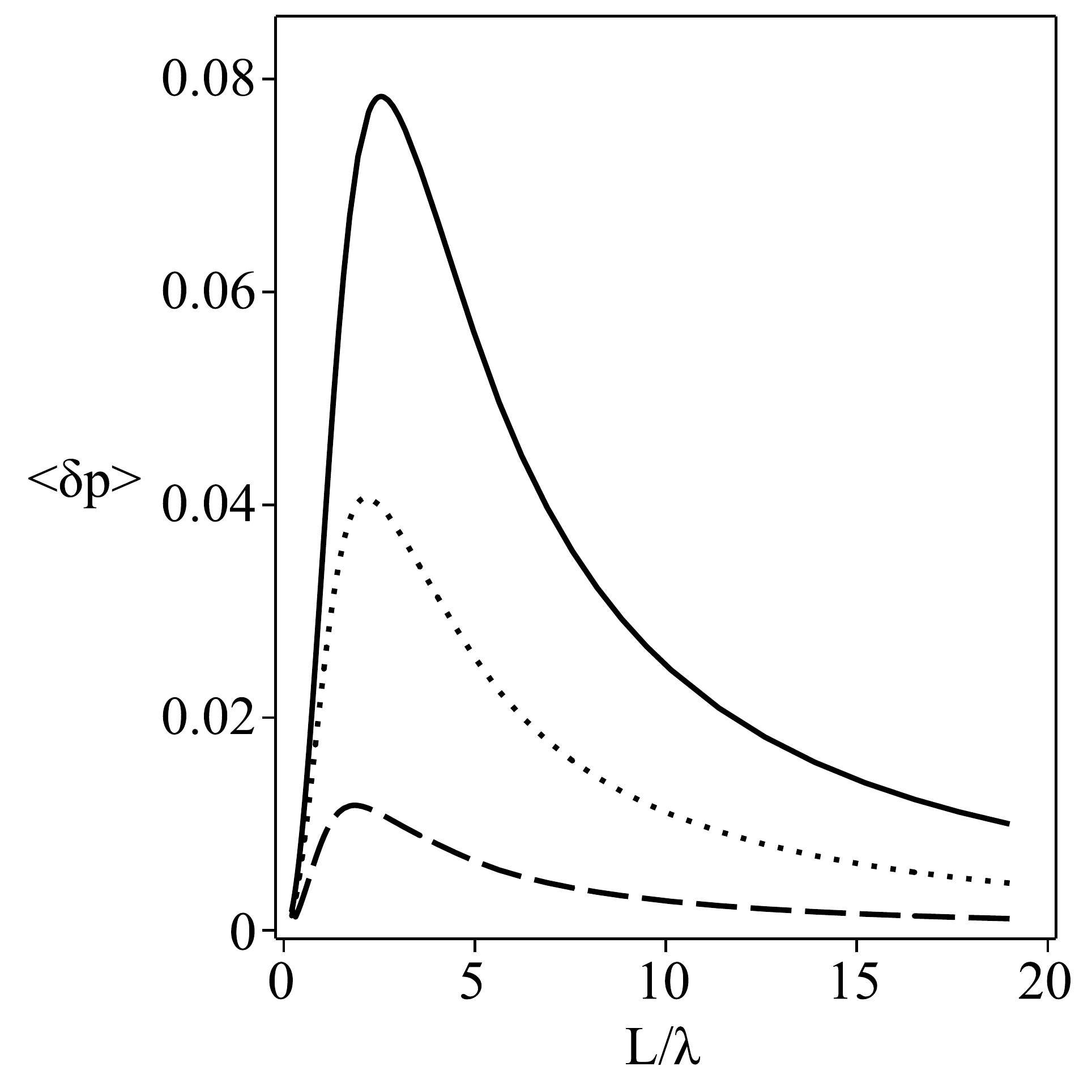}
\end{center}
\caption{Excess mean polarization $\langle\delta p\rangle$ as a function of $L/\lambda\sim\sqrt{E}$ obtained from averaging the numerical solutions of Eqs. \eqref{sigma-p2} and \eqref{px2} for three different values of activity $\nu=\nu_0=\nu_\alpha=\nu_\beta$ : $\nu=0.5$ (dashed line), $\nu=1.0$ (dotted line) and $\nu=1.5$ (solid line). The plots are for $\tilde{w}=4$, $\tilde{a}=\tilde{b}=1$ and $\tilde{K}=1$.}
\label{polarization}
\end{figure}

To quantify the different properties of these two states, we define an excess mean polarization averaged over the cell as $\langle\delta p\rangle=\int_0^L\frac{dx}{L}[p(x)-p_0]$.  The excess polarization $\langle\delta p\rangle$ is zero for the symmetric polarization profiles obtained with the boundary condition $p(0)=p(L)=0$, whereas $\langle\delta p\rangle$ obtained for the boundary condition $p(0)=p(L)=\sqrt{-a/b}$ is a non-monotonic function of substrate stiffness, as shown in Fig. \ref{polarization} for three values of activity. The excess polarization is largest at a characteristic substrate stiffness, suggesting that enhancement of stress fiber and resulting cell polarization may be obtained for an optimal substrate rigidity, as reported in ~\cite{Zemel2010}.  The  excess polarization $\langle\delta p\rangle$ vanishes in the absence of activity and its maximum value increases with activity.

We have presented a minimal continuum model of the interaction of a cell adhering to an elastic substrate. The cell is described as an active elastic gel and the coupling to the substrate enters as a boundary condition. The model shows that the interplay of substrate coupling and activity yields contractile stresses and deformation in the cell and can enhance polarization, breaking the front/rear symmetry of the cell. The model provides a simple, yet powerful continuum formulation for the description of cell-substrate interactions and can be extended in various directions by considering more realistic two-dimensional cell geometries and anisotropic and deformable substrates. The possibility of cell migration will also be incorporated in future work. Finally, the continuum model can be used to describe the interaction of confluent layers of epithelial cells with substrates. In this case a direct comparison with recent experiments that have imaged the stress distribution in migrating cell layers~\cite{Tambe2011} may be possible.

\acknowledgments
{\bf Acknowledgements}
This work was supported by the National Science Foundation through awards DMR-0806511 and NSF-DMR-1004789.  We thank Yaouen Fily and Silke Henkes for important feedback on cell-substrate elasticity and Tannie Liverpool for illuminating discussions on active systems in general.

\end{document}